\documentclass[aps,prl,twocolumn,superscriptaddress]{revtex4-1}
\usepackage[compat=1.0.0]{tikz-feynman}

\usepackage{amssymb}
\usepackage{multirow}
\usepackage{bm}
\usepackage{amsmath}
\usepackage{graphicx}
\usepackage{epstopdf}
\usepackage{subfigure}
\usepackage{natbib}
\usepackage{epsfig}
\usepackage{amsfonts}
\usepackage{mathrsfs}
\usepackage[toc,page,title,titletoc,header]{appendix}
\usepackage[colorlinks,linkcolor=blue,citecolor=blue,anchorcolor=blue]{hyperref}

\usepackage{dsfont,amsthm,amsbsy}

\usepackage{fancyhdr}

\usepackage{ulem}

\begin{document}
\title{Pair-density-wave and $s\pm \mathrm{i}d$ superconductivity \\
in a strongly coupled, lightly doped Kondo insulator}
\author{Fangze~Liu}
\affiliation{Department of Physics, Stanford University, Stanford, California 94305, USA}
\affiliation{Stanford Institute for Materials and Energy Sciences, SLAC National Accelerator Laboratory, 2575 Sand Hill Road, Menlo Park, CA 94025, USA}
\author{Zhaoyu~Han}
\email{zyhan@stanford.edu}
\affiliation{Department of Physics, Stanford University, Stanford, California 94305, USA}

\begin{abstract}
We investigate the large Kondo coupling limit of the Kondo-Heisenberg model on one- and two-dimensional lattices. Focusing on the possible superconducting states when slightly doping the Kondo insulator state, we identify different pairing modes to be most stable in different parameter regimes. Possibilities include uniform $s$-wave, pair-density-wave with momentum $\pi$ (in both one and two dimensions) and uniform $s\pm \mathrm{i}d_{x^2-y^2}$-wave (in two dimensions). We attribute these exotic pairing states to the presence of various pair-hopping terms with a ``wrong'' sign in the effective model, a mechanism that is likely universal for inducing pairing states with spatially modulated pair wavefunctions.

\end{abstract}

\maketitle

The Kondo-Heisenberg model is a paradigmatic model of strongly correlated electronic systems, attracting interest as a model of various materials~\cite{coleman2015heavy, RevModPhys.69.809,gulacsi2006kondo,PhysRevB.101.020501, PhysRevB.106.045103,gall2021competing,sompet2022realizing,PhysRevLett.129.047601,PhysRevLett.131.026501,PhysRevLett.131.026502,PhysRevB.107.245102} and as a prototypical model for various exotic physical phenomena including quantum criticality~\cite{DONIACH1977231,doi:10.1126/science.1191195,PhysRevLett.98.026402,si2014kondo,PhysRevLett.122.217001}, fractionalization~\cite{PhysRevLett.90.216403, PhysRevB.69.035111}, odd frequency pairing~\cite{PhysRevB.49.8955,doi:10.1073/pnas.1902928116} and pair density wave~\cite{PhysRevB.109.014103,doi:10.1073/pnas.1902928116,PhysRevB.63.205104,PhysRevLett.105.146403,PhysRevB.101.165133}. Although it has received considerable theoretical investigations, the majority of research efforts have adopted bosonization techniques in one dimension~\cite{RevModPhys.69.809,gulacsi2006kondo,gulacsi2004one,PhysRevB.64.033103,PhysRevB.63.205104,PhysRevB.59.15641} or large-$N$ techniques~\cite{read1983solution,PhysRevB.29.3035,PhysRevLett.66.1773} in two or higher dimensions~\cite{coleman2015heavy,coleman1989kondo,PhysRevB.102.115133,PhysRevB.92.094401,PhysRevLett.100.236403,PhysRevB.56.11820,PhysRevLett.105.246404,read1983new,PhysRevLett.57.877,PhysRevB.35.5072,PhysRevB.75.165110,zhong2015fermionology,PhysRevB.30.3841,liu2014pairing}, and, with a few exceptions~\cite{lacroix1985some,PhysRevB.46.13838,RevModPhys.69.809,gulacsi2006kondo,bastide1987d,coleman2010frustration}, focused on relatively weak coupling regimes (\textit{i.e.}, the Kondo coupling strength is not strong compared to the bandwidth of the itinerant electrons).

\begin{figure*}[htb!]
\includegraphics[width=1\linewidth]{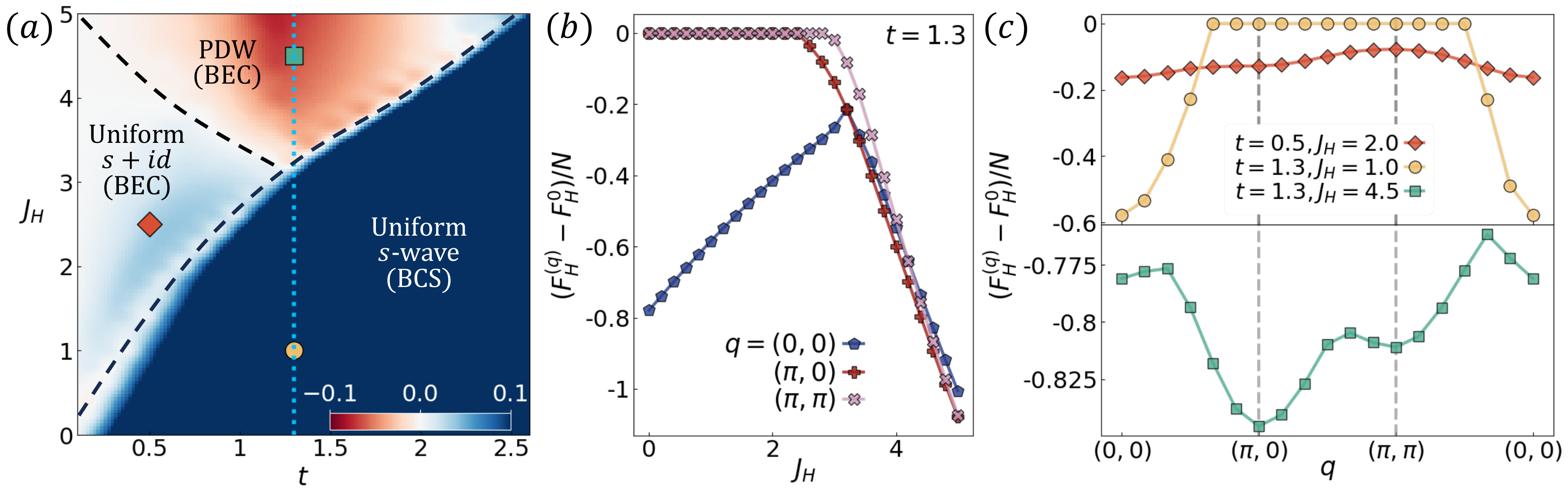}
\caption{\label{2D_F} \textbf{Condensation energy in 2D square lattice.} (\textbf{a}) Condensation energy difference (per site) between $\bm{q}=(\pi,0)$ and $\bm{q}=(0,0)$, \textit{i.e.}, $(F_\text{H}^{(\pi,0)} - F_\text{H}^{(0,0)})/N$. The black dashed lines serve as visual guides to demarcate distinct phase regions. (\textbf{b}) Examples of condensation energies at several high-symmetry $\bm{q}$'s along the vertical dotted line cut in (a). (\textbf{c}) The full $\bm{q}$ dependence for a few representative choices of parameters for different phases marked in (a). Simulations are done with a $N=12\times 12$ lattice for (a) and (b), and a $N=24\times 24$ lattice for (c).  }
\end{figure*}

\begin{figure*}
\includegraphics[width=1\linewidth]{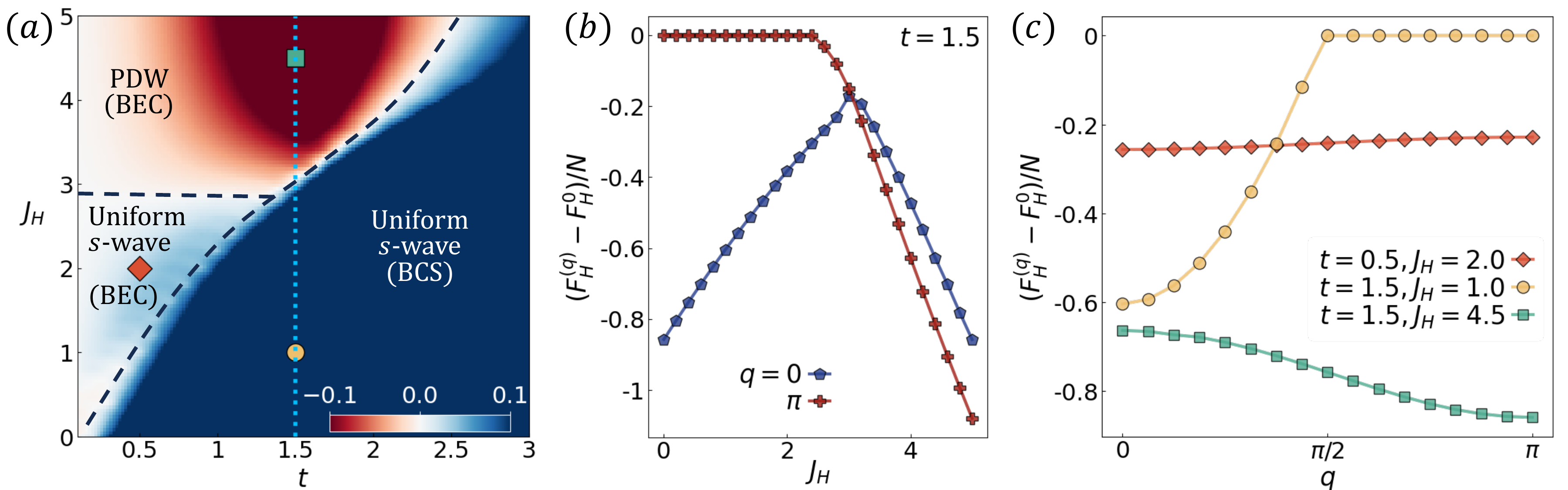}
\caption{\label{1D_F} \textbf{Condensation energy in 1D lattice.} (\textbf{a}) Condensation energy difference (per site) between $q=\pi$ and $q=0$, \textit{i.e.}, $(F_\text{H}^{(\pi)} - F_\text{H}^{(0)})/N$. The black dashed lines serve as visual guides to demarcate distinct phase regions. (\textbf{b}) Examples of condensation energies at several high-symmetry $\bm{q}$'s along the vertical dotted line cut in (a). (\textbf{c}) The full $\bm{q}$ dependence for a few representative choices of parameters for different phases marked in (a). Simulations are done with a $N=128$ chain for (a) and (b), and a $N=256$ chain for (c).}
\end{figure*}

In pursuit of an in-depth understanding of this important model, this work focuses on the large Kondo coupling regime, where the (antiferromagnetic) Kondo coupling $J_\text{K}$ is much greater than the electron hopping amplitude $|t|$ and the local moment (antiferromagnetic) Heisenberg coupling $J_\text{H}$.
In this limit, a mapping to the infinite $U$ Hubbard model~\cite{lacroix1985some} and a strong coupling expansion can be justified, based on which we explicitly derive the low-energy effective model. This model is similar to the $t-J$ model derived from the strong coupling limit of the Hubbard model but features various additional pair-hopping terms. Based on this effective model, we investigate the superconducting (SC) phase diagram at a filling fraction slightly away from one electron per unit cell and at a low temperature, by means of numerical mean-field (MF) theory. For each set of parameters ($t/J_\text{K}$ and $J_\text{H}/J_\text{K}$), we solve the pairing wavefunction self-consistently and compute the free energy at every Cooper pair momentum $\bm{q}$, allowing us to determine the optimal pairing momentum, and, if the pairing is at certain high-symmetry momentum, the pairing symmetry. The phase diagrams are summarized for the 2D and 1D cases in Figs.~\ref{2D_F}\&~\ref{1D_F}, respectively. Remarkably, in the 2D scenario, an $s\pm \mathrm{i} d_{x^2-y^2}$ pairing state featuring broken time-reversal symmetry is found to be stable within a regime of the phase diagram. More interestingly, within a similar parameter regime for both cases, a pair-density-wave (PDW)~\cite{agterberg2020physics} with momentum $\pi$ is found to be the most stable state, in possible agreement with the 1D result obtained by a bosonization method~\cite{PhysRevB.63.205104} and numerical density-matrix-renormalization-group studies~\cite{PhysRevLett.105.146403,PhysRevB.101.165133}. We explain the observed exotic pairing phases from a strong-pairing perspective based on the observation that the leading pair-hopping terms in the effective model have a ``wrong'' sign, which is likely a general mechanism for such exotic pairing momenta and/or symmetries (\textit{e.g.}, Refs.~\cite{PhysRevLett.125.167001,PhysRevLett.130.026001}, for similar scenarios).

{\bf Model and Method. } In this paper, we study the Kondo-Heisenberg model:
\begin{align}
    \hat{H} =& -t\sum_{\langle ij \rangle,\sigma }(\hat{c}^{\dagger}_{i\sigma} \hat{c}_{j\sigma}  +\text{h.c.}) + J_\text{K} \sum_{i} \hat{\bm{s}}_i \cdot \hat{\bm{S}}_i \nonumber\\ 
    & \ \ + J_\text{H} \sum_{\langle ij \rangle} \hat{\bm{S}}_i \cdot \hat{\bm{S}}_j
\end{align}
where $\hat{c}_{i\sigma}$ annihilates a spin-$\sigma$, itinerant electron on site-$i$, and $\hat{\bm{s}}_i $ and $\hat{\bm{S}}_i$ respectively represent the electron spin and the local moment on site $i$. While this model is definable on any lattice, for concreteness we will focus on the 1D chain and the 2D square lattice. Due to the bipartite nature of the lattices, there is a particle-hole symmetry generated by $\hat{c}_{i\sigma} \rightarrow (-1)^i \hat{c}^\dagger_{i\sigma}$, allowing us to concentrate on the hole-doped side, where $n \equiv \frac{1}{N} \sum_{i} \langle \hat{c}^\dagger_{i\sigma}\hat{c}_{i\sigma} \rangle  < 1$ ($N$ is the system size). Furthermore, since the sign of $t$ can be trivially altered by a gauge transformation $\hat{c}_{i\sigma} \rightarrow (-1)^i \hat{c}_{i\sigma}$, we assume $t>0$ without loss of generality.

In this work, we consider the large $J_\text{K}$ limit by regarding $t/J_\text{K}$ and $J_\text{H}/J_\text{K}$ as small parameters. To the zeroth order of the analysis and when electron filling $n< 1$, each site has three possible states: $(|\Uparrow \downarrow\rangle - |\Downarrow \uparrow\rangle)/\sqrt{2}$ (Kondo singlet) or $| \varnothing \updownarrow\rangle $, where $\Updownarrow$ represents the spin of the electron whereas $\updownarrow$ represents the local moment, and $\varnothing$ indicates the absence of any itinerant electron. Different tensor-product combinations of these states form the low-energy Hilbert space, $\mathcal{H}_\text{eff}$, which is separated from all the other states by an energy gap $\sim J_\text{K}$. To describe the low-energy physics within $\mathcal{H}_\text{eff}$, we define a set of fermionic operators $\hat{h}_{i\sigma}$ to effectively describe the holes doped into the system. The mapping between these hole operators and the operators in  $\mathcal{H}_\text{eff}$ can be locally established as~\cite{lacroix1985some}
\begin{align}
\hat{h}_{i\updownarrow} \leftrightarrow \frac{1}{\sqrt{2}}(|\Uparrow \downarrow\rangle - |\Downarrow \uparrow\rangle) \langle  \varnothing \updownarrow |, 
\end{align}
thus $\hat{h}_{i\sigma}$ annihilates a spin-$\sigma$, charge-$-1$ object relative to the `vacuum' of $\mathcal{H}_\text{eff}$, which refers to the strong coupling Kondo insulator state with a Kondo singlet on every site. Note that, to faithfully map between in the physical Hilbert space $\mathcal{H}_\text{eff}$ and the Fock space of the hole operators, it needs to be further recognized that two holes cannot simultaneously occupy the same site.

We then perform a perturbation expansion to derive a low-energy effective Hamiltonian. 
Leveraging the above equivalence mapping of Hilbert space, we express the result in terms of the hole operators, including all terms to the zeroth and the first order in powers of $1/J_{\text{K}}$:
\begin{align} \label{eq: eff H}
H_\text{eff} =& \hat{P}(\hat{H}_t + \hat{H}_\tau + \hat{H}_V)\hat{P} \\
\hat{H}_t =& t_1 \sum_{\langle ij \rangle \sigma}  \left(\hat{h}_{i\sigma}^{\dagger} \hat{h}_{j \sigma}  + \text{h.c.}\right)\nonumber\\
& + t_2 \sum_{\langle ijk \rangle \sigma} \left(\hat{h}_{i\sigma}^{\dagger} \hat{h}_{k \sigma}  + \text{h.c.}\right) \label{eq: Ht} \\
\hat{H}_V = &J_{\text{H}}\sum_{\langle ij \rangle} \hat{\bm{S}}_i \cdot \hat{\bm{S}}_j +V \sum_{\langle ij \rangle} \hat{n}_i^h \hat{n}_j^h  \nonumber\\
&-J' \sum_{\langle ijk\rangle} \hat{\bm{S}}_i\cdot\hat{\bm{S}}_k(1-\hat{n}_j^h) \label{eq: HV} \\ 
\hat{H}_\tau = &\sum_{\langle ijk\rangle} \Bigg{[}  t_1' \sum_{\sigma} \hat{h}^\dagger_{i\sigma} \hat{n}^h_{k} \hat{h}_{j\sigma} + \tau_1 \hat{\xi}^\dagger_{ik}\hat{\xi}_{kj} + (i \leftrightarrow k)  \nonumber\\
&\ \ \ \ \ \ \  +  t_2'\sum_{\sigma} \hat{h}^\dagger_{i\sigma} \hat{n}^h_{j} \hat{h}_{k\sigma}  + \tau_2\hat{\xi}^\dagger_{ij}\hat{\xi}_{jk} +\text{h.c.} \Bigg{]} \label{eq: Htau} 
\end{align}
where $\langle ijk\rangle$ represents a triplet of sites in which site $j$ is a nearest-neighbor of two distinct sites $i$ and $k$, $\hat{n}^h_i \equiv \sum_{\sigma} \hat{h}^\dagger_{i\sigma} \hat{h}_{i\sigma} $ is the hole density on site-$i$, and $\hat{P}$ is a projector enforcing the Hilbert constraint, \textit{i.e.}, excluding the states with double occupation of holes on any site. The effective parameters are $t_1  = \frac{t}{2} - \frac{3 t J_{\text{H}}}{4 J_{\text{K}}}$, $t_2 = \frac{t^2}{6 J_{\text{K}}}$, $V = (\frac{5t^2}{6J_{\text{K}}}+\frac{9J_{\text{H}}^2}{32J_{\text{K}}})$, $J' = \frac{J_{\text{H}}^2}{2 J_{\text{K}}}$, $t_1' = - \frac{t J_{\text{H}}}{8 J_{\text{K}}}$,  $\tau_1 = \frac{t J_{\text{H}}}{2J_{\text{K}}}$, $t_2' = \frac{t^2}{12J_{\text{K}}} $, and $\tau_2 = \frac{t^2}{2J_{\text{K}}} $. For convenience, we have defined $\hat{\xi}_{ij} \equiv (\hat{h}_{i\uparrow} \hat{h}_{j\downarrow} + \hat{h}_{j\uparrow} \hat{h}_{i\downarrow})/\sqrt{2}$, the singlet annihilation operator on sites $i$ and $j$. We note that a similar mapping and expansion have been done for the Kondo lattice model without Heisenberg coupling~\cite{RevModPhys.69.809,gulacsi2006kondo,bastide1987d}, and our results agree with the existing literature upon setting $J_\text{H} = 0$. We also note that the generalization of this expansion to other similar physical models (\textit{e.g.}, the one in Ref.~\cite{PhysRevB.101.020501}) can be done in a systematic manner.

To mitigate the complexities of this Hamiltonian, we invoke an exact rewriting (for any $i,j$)
\begin{align}
    \hat{P}\hat{h}^{\dagger}_{i\sigma} \hat{h}_{j\sigma}\hat{P} = (1-\hat{n}^h_{i\bar{\sigma}})\hat{h}^{\dagger}_{i\sigma} \hat{h}_{j\sigma} (1-\hat{n}^h_{j\bar{\sigma}})
\end{align}
to equivalently implement the projection. Then, we consider the dilute hole limit, \textit{i.e.}, $n^h\equiv 1 - n \ll 1$. In this limit, the expectation values of all pairs of fermion operators, \textit{i.e.}, $\hat{h}^{\dagger}\hat{h}$, are bounded by $n^h$. Therefore, we neglect the terms consisting of more than four fermion operators, as they are of order $\mathcal{O}[(n^h)^3]$ and thus insignificant relative to other terms. 

After these manipulations of the Hamiltonian, writing in momentum space, we obtain a standard interacting Hamiltonian for fermions:
\begin{align}
    H_\text{eff} \approx & \sum_{\bm{k}\sigma} \epsilon_{\bm{k}} \hat{h}^\dagger_{\bm{k} \sigma} \hat{h}_{\bm{k} \sigma} \nonumber\\
    &+ \frac{1}{N} \sum_{\substack{\bm{k},\bm{k}';\\
    \sigma, \sigma';\bm{q}}} \Gamma^{\sigma, \sigma'}_{\bm{k},\bm{k}';\bm{q}} \hat{h}^\dagger_{\frac{\bm{q}}{2}-\bm{k} \sigma}\hat{h}^\dagger_{\frac{\bm{q}}{2}+\bm{k}  \sigma'} \hat{h}_{\frac{\bm{q}}{2}+\bm{k}'  \sigma'} \hat{h}_{\frac{\bm{q}}{2}-\bm{k}'  \sigma}
\end{align}
where the expressions of the bare dispersion $\epsilon_{\bm{k}} $ and interacting vertex $\Gamma^{\sigma, \sigma'}_{\bm{k},\bm{k}';\bm{q}}$ are explicitly given in Supplementary Materials (SM)~\footnote{See [url] for the explicit expressions of the effective model, the detailed discussions about the mean field calculation and the results. }.

Finally, we perform MF analysis for the possible SC states in the system. Due to the spin rotation symmetry, we can focus on the sector with $\sigma'=\bar{\sigma}$, since the other two triplet states are degenerate with the one with $S^z =0$. Then for each possible Cooper pair momentum $\bm{q}$, we adopt the Bogoliubov-de-Gennes MF ansatz (note that in the equation below, $\bm{q}$ is no longer a dummy variable)
\begin{align}
    H^{(\bm{q})}_{\text{MF}} = & \sum_{\bm{k}\sigma} \epsilon_{\bm{k}} \hat{h}^\dagger_{\bm{k} \sigma} \hat{h}_{\bm{k} \sigma} \nonumber\\
    &+ \frac{2}{N} \sum_{\bm{k},\bm{k}'} \Gamma^{\uparrow \downarrow}_{\bm{k},\bm{k}';\bm{q}} \left[ \hat{h}^\dagger_{\frac{\bm{q}}{2}-\bm{k} \downarrow}\hat{h}^\dagger_{\frac{\bm{q}}{2}+\bm{k} \uparrow} \phi_{\bm{k}'}^{(\bm{q})} +\text{h.c.}\right]
\end{align}
where 
\begin{align} \label{eq: MF eq}
\phi^{(\bm{q})}_{\bm{k}'}\equiv \langle \hat{h}_{\frac{\bm{q}}{2}+\bm{k}'  \uparrow} \hat{h}_{\frac{\bm{q}}{2}+\bm{k}'  \downarrow}\rangle_{\text{MF}}
\end{align}
gives the MF self-consistency equation that we will solve by numerical iteration.

{\bf Numerical results. } For each set of parameters $\{t/J_{\text{K}}, J_{\text{H}}/J_{\text{K}},n^h\}$, we perform MF calculation on various different $\bm{q}$ values. For each specific $\bm{q}$, we solve the MF equation Eq.~\ref{eq: MF eq} and obtain a solution with the lowest Helmholtz free energy, $F^{(\bm{q})}_{\text{H}}$. By comparing the free energies for different $\bm{q}$ values, the optimal SC states can then be determined. The condensation energy can be further obtained by comparing the free energy $F^{(\bm{q})}_{\text{H}}$ with that of a system without SC order, denoted as $F^0_\text{H}$.

For all the computations presented in this study, we set $J_\text{K}=10$ as the large {\it energy scale} and explore the system's behavior at $T = 1/20$, the lowest temperature at which we can attain well-converged solutions~\footnote{Due to the finite sizes employed in our study, which result in a finite resolution of the density of states, fixing the density becomes challenging at lower temperatures. }. For these calculations, we properly choose the chemical potential to ensure a hole density of $n^h = 1/8$.  We systematically explore a range of relatively small values for $J_\text{H}\in [0,5]$ and $t\in[0,3]$ are studied. The primary results are presented in the main text, while more detailed data can be accessed in the SM~\cite{Note1}.

We first investigate the 2D square lattice that is of most interest. The results of the condensation energy density are summarized in Fig.~\ref{2D_F}. In Fig.~\ref{2D_F}a\&b, it is evident that over a broad region at large $J_\text{H}$, a PDW state with Cooper pair momentum $\bm{q} = (\pi,0)$ (or $(0,\pi)$) is energetically more favorable, and we have verified that there is no other competing $\bm{q}$ within the entire Brillouin zone. To provide a better illustration, we select three representative sets of parameters and plot their condensation energy density as a function of $\bm{q}$ in Fig.~\ref{2D_F}c. For $(t,J_\text{H})=(1.3,4.5)$, it is clear that the energy minimum is located at $(\pi,0)$ (or $(0,\pi)$). It should also be noted that the three curves have notable qualitative distinctions. Clearly, two of them (with $(t,J_\text{H})=(1.3,4.5)$ and $(0.5,2.0)$) have a small curvature around the minimum and a narrow bandwidth relative to the condensation energy, whereas the other point (with $(t,J_\text{H})=(1.3,1.0)$) has the opposite features. This observation suggests that the $J_\text{H}\gtrsim 2t$ region is in a strong pairing regime that can be more suitably described by a Bose-Einstein Condensation (BEC) of preformed pairs, whereas the $J_\text{H}\lesssim 2t$ region is a weak pairing region and can be effectively described by the Bardeen-Cooper-Schrieffer (BCS) theory.

\begin{figure}
\includegraphics[width=1\linewidth]{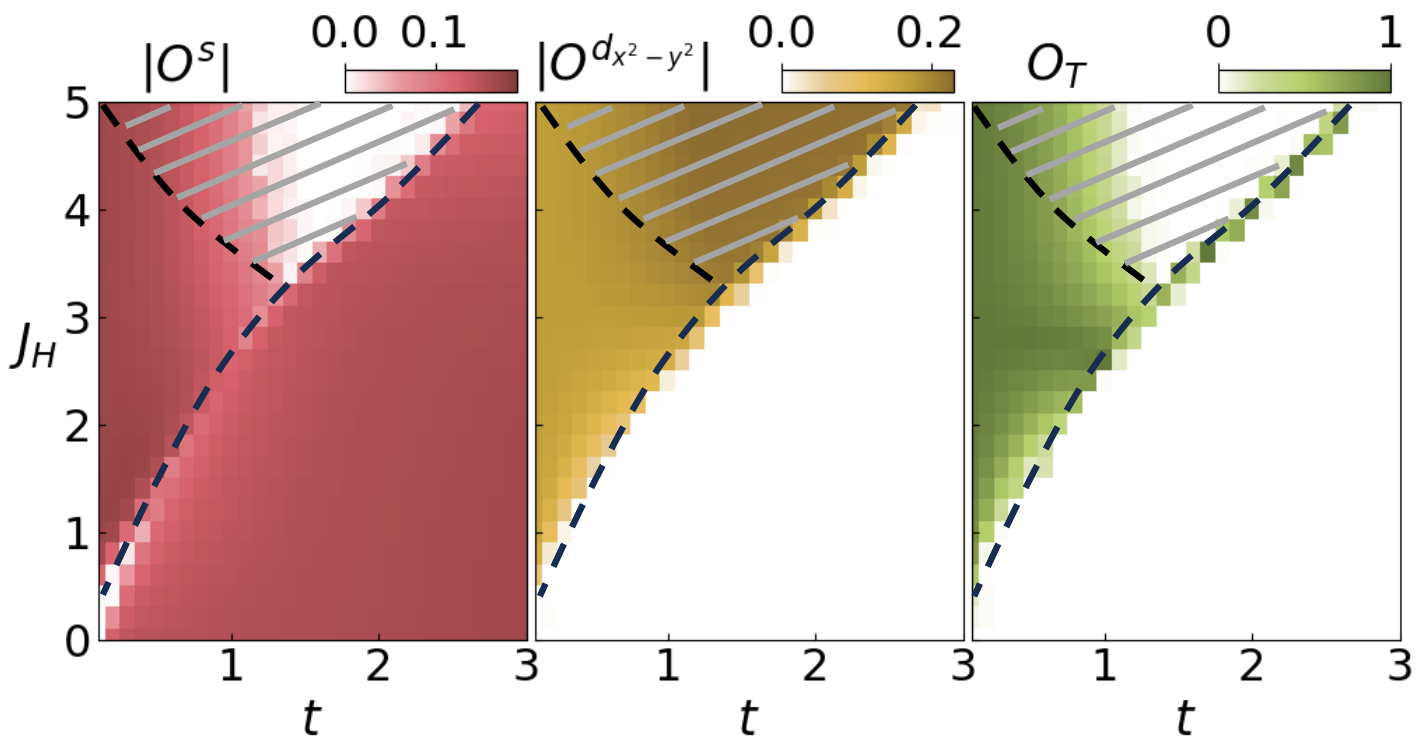}
\caption{\label{2D_sym} 
The magnitude of pairing order parameters, as defined in Eqs.~\ref{eq: order parameter}\&\ref{eq: T order parameter}, for the uniform pairing states ($\bm{q} = (0,0)$) in 2D. Simulations are done on a $12\times 12$ lattice. Note that as in Fig.~\ref{2D_F}a, distinct phase regions are delineated by black dashed lines; within the dashed-out region, the uniform pairing state is less favorable than the PDW state.}
\end{figure}

\begin{figure}
\includegraphics[width=1\linewidth]{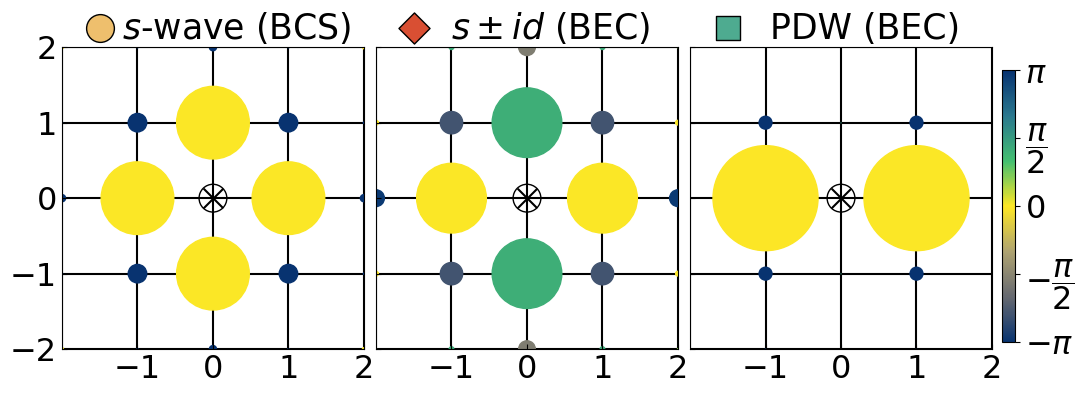}
\caption{\label{2D_phi} 
The magnitude (represented by circle radius) and phase (indicated by color) of the normalized pair field $\phi_{\bm{r}}^{(\bm{q})}/\sqrt{\sum_{\bm{r}} |\phi_{\bm{r}}^{(\bm{q})}|^2}$ (\textit{i.e.}, the wavefunction of the Cooper pairs) as a function of $\bm{r}$ (the relative coordinate between the two electrons in a cooper pair), at several representative parameter points marked in Fig.~\ref{2D_F}a. The $\phi^{(\bm{q})}_{\bm{r} = (0,0)}$ is manually set to zero in these plots to respect the Hilbert space constraint. The parameters for these representative points are as follows: 
(left) $t=1.3, J_\text{H}=1, \bm{q}=(0,0)$; 
(middle) $t=0.5, J_\text{H}=2, \bm{q}=(0,0)$; 
(right) $t=1.3, J_\text{H}=4.5, \bm{q}=(\pi,0)$. Simulations are done on a $24\times 24$ 2D square lattice. }
\end{figure}

To further investigate the pairing symmetries of the uniform pairing ($\bm{q}=(0,0)$) states occupying most parts of the phase diagram in Fig.~\ref{2D_F}a, we compute several order parameters defined as:
\begin{align} \label{eq: order parameter}
O^{\ell } \equiv \frac{1}{N} \sum_{\bm{k}} f^{\ell}_{\bm{k}} \phi^{(0,0)}_{\bm{k}}
\end{align}
where $\ell = s,d_{xy},d_{x^2-y^2},(p_x,p_y)$ are the irreducible representations of the $D_4$ group, and $f^{\ell}$ are the corresponding form factors. For the pairing symmetries that are non-zero in our case, we take $f^{s}_{\bm{k}} \equiv \cos{k_x}+\cos{k_y}$, and $f^{d_{x^2-y^2}}_{\bm{k}} \equiv \cos{k_x}-\cos{k_y}$. To detect time-reversal symmetry breaking, we compute
\begin{align} \label{eq: T order parameter}
    O_{\mathcal{T}} \equiv 1 - \frac{|\sum_{\bm{k}}(\phi^{(0,0)}_{\bm{k}})^2|}{\sum_{\bm{k}}|\phi^{(0,0)}_{\bm{k}}|^2}
\end{align}
The amplitudes of these order parameters are plotted in Fig.~\ref{2D_sym}. It is probably not surprising to see that the BCS uniform pairing state is a pure $s$-wave state. However, interestingly, we find the BEC uniform phase has coexisting $s$ and $d_{x^2-y^2}$ pairing components, and the time-reversal symmetry is also spontaneously broken, suggesting an exotic $s\pm \mathrm{i}d_{x^2-y^2} $ pairing. This finding gains further support through the direct visualization of the pairing wavefunctions in real space in Fig.~\ref{2D_phi}, where it can be directly seen that the relative phase between the pair fields on the nearest neighbor bonds in $x$ and $y$ directions is $\pi/2$. It is also remarkable that in the dashed-out regime where the uniform pairing state gives way to the PDW state, the uniform pairing state itself crossovers from an $s+\mathrm{i} d_{x^2-y^2}$ to a $d_{x^2-y^2}$-wave state, and has competitive energy compared to the PDW state (Fig.~\ref{2D_F}c). We hope that our MF results can serve as an invitation for further investigations of the problem in this relatively unexplored parameter regime. 

Although mean-field theories are generically less reliable in 1D due to strong fluctuations, we nonetheless performed the same analysis for the 1D chain case, with the aim of facilitating comparison with existing results. The outcomes, as depicted in Fig.~\ref{1D_F}, closely resemble the findings in the 2D scenario, and the differences compared to the 2D case are 1) the PDW state with $\bm{q}=\pi$ is favorable in an even broader regime, and 2) the BEC uniform pairing state is no longer exotic. It is encouraging to note that, density-matrix-renormalization-group (DMRG) studies have found PDW to be the ground state at $J_\text{H}/J_{\text{K}} = 1$, $t/J_{\text{K}}=1/2$~\cite{PhysRevLett.105.146403,PhysRevB.101.165133}, a point that is possibly connected to the PDW regime in Fig.~\ref{1D_F}a after extrapolating the phase boundary. 

{\bf A possible mechanism of the exotic SC states. } As seen in Fig.~\ref{2D_phi}, we find that the interesting PDW and $s \pm \mathrm{i} d_{x^2-y^2}$ states have dominant pairing amplitude on the nearest neighbor bonds. On the other hand, from Fig.~\ref{2D_F}c it can be seen that the energy gain associated with the pair formation $\sim |F^{(\bm{q})}_\text{H}/(n^hN)|$ (or the single-particle gap presented in SM~\cite{Note1}), is much higher than the phase stiffness $\sim\nabla^2_{\bm{q}} F^{(\bm{q})}_\text{H}/N$ at the optimal $\bm{q}$, so it can be concluded that these pairs are pre-formed before phase coherence develops. This can be intuitively understood by the presence of a strong $J_\text{H}$ which stabilizes such a local singlet pairing at a relatively high energy scale. This observation motivates us to take a perspective starting from these preformed ``bond dimers'' by considering an effective dimer theory (subject to hard-core constraints that are relatively unimportant in the dilute limit due to the low collision probability):
\begin{eqnarray}
\hat{H}_\mathrm{dimer}=-\sum_{\langle ij\rangle,\langle mn\rangle} \left( \tau_{ij,mn} \hat \xi^\dag_{ij} \hat \xi_{mn} +\textrm{h.c.}\right),
\label{dimer}
\end{eqnarray}
where $\tau_{ij,mn}$ is the effective pair hopping amplitude between bond $\langle ij\rangle$ and bond $\langle mn\rangle$. The BEC of these dimers onto the boson band minima determines the pairing mode of the SC state. From the form of the effective Hamiltonian in Eq.~\ref{eq: eff H}, the leading terms that can contribute to the boson hopping matrix are the $t_2$ terms in Eq.~\ref{eq: Ht} and $t_2'$, $\tau_2$ terms in Eq.~\ref{eq: Htau}, which can move a bond dimer to another bond with a shared site. The crucial thing that allows the exotic pairing states to arise in this system is that these dominant terms contribute {\it negatively} to the hopping matrix $\tau_{ ij, mn}$, circumventing the limitation of the Perron-Frobenius theorem that always gives rise to a uniform $s$-wave pairing ground state and is applicable when all matrix entries are non-negative. Actually, these leading contributions yield an exactly flat boson band at low energy, which opens room for the higher-order perturbations in the boson hopping matrix to lift the degeneracy and lead to an exotic pairing momentum and/or symmetry. This picture for PDW based on the ``wrong'' signs of certain pair-hopping terms seems to be a general, strong coupling mechanism already seen to be valid in several systems~\cite{PhysRevLett.125.167001,PhysRevLett.130.026001} and may be responsible for the PDW phase in other systems~\cite{PhysRevB.106.045103, PhysRevB.108.035135}.

{\bf Acknowledgement. } We thank helpful discussions with Steven Kivelson and Srinivas Raghu. This work is funded by the Department of Energy, Office of Basic Energy Sciences, Division of Materials Sciences and Engineering, under contract DE-AC02-76SF00515 at Stanford.

\bibliography{ref}

\end{document}

\title{\Large Supplementary Material\\
\large Pair-density-wave and $s+\mathrm{i}d$ superconductivity \\
in a strongly coupled, lightly doped Kondo insulator}
\author{Fangze~Liu}
\affiliation{Department of Physics, Stanford University, Stanford, California 94305, USA}
\author{Zhaoyu~Han}
\affiliation{Department of Physics, Stanford University, Stanford, California 94305, USA}

\maketitle

In this Supplementary Material, we provide a comprehensive explanation of our calculation covering (i) the strong-coupling expansion, (ii) the effective Hamiltonian, (iii) the mean-field calculations, (iv) the symmetries and the order parameters under consideration, and (v) illustrative simulations of singlet superconducting gap functions obtained from our mean-field calculations.

\section{Strong-coupling expansion}

We initiate our analysis with Kondo coupling

\begin{equation}
\begin{aligned}
\hat{H}_{0} = J_K \sum_{j} \bm{\hat{s}}_j \cdot \bm{\hat{S}}_j
= J_K \sum_{j} \big( \frac{1}{2}(\hat{s}_j^{+} \hat{S}_j^{-}+\hat{s}_j^{-}\hat{S}_j^{+}) + \hat{s}_j^{z} \hat{S}_j^{z} \big),
\end{aligned}
\end{equation}
Here, $\bm{\hat{S}}_j$ represents the spin-$\frac{1}{2}$ operator in the Heisenberg layer, whereas $\bm{\hat{s}}_j$ corresponds to the conduction layer.\\

The eigenstates and corresponding eigenenergies are
\begin{center}
\begin{tabular}{ c || c | c | c }
 \hline
 Eigenstates & $\frac{1}{\sqrt{2}} (\mtrxud{\Uparrow}{\downarrow} - \mtrxud{\Downarrow}{\uparrow})=(\Diamond)$ & $\mtrxud{\varnothing}{S}$, $\mtrxud{\Uparrow\Downarrow}{S} $ & $\frac{1}{\sqrt{2}} (\mtrxud{\Uparrow}{\downarrow} + \mtrxud{\Downarrow}{\uparrow})=(\vrec)$, $\mtrxud{s}{S}$\\
 \hline
 Eigenvalues & $-\frac{3}{4}J_K$ & 0 &  $\frac{1}{4}J_K$\\
 \hline
\end{tabular}
\end{center}
where $s=\{\Uparrow,\Downarrow\}$ represents the spin of the electron whereas $S=\{\uparrow,\downarrow\}$ represents the local moment, and $\varnothing$ indicates the absence of an electron.

The ground state of the doped $\hat{H}_0$ is
\begin{equation}
\begin{aligned}
\Psi_0 = \prod_{i\in \text{holes}} \mtrxud{\varnothing}{S}_i \prod_{j\in \text{electrons}}(\Diamond)_j
\end{aligned}
\end{equation}

We consider perturbations in the strong $J_K$ case:
\begin{equation}
\begin{aligned}
\hat{V}_1 &= -t \sum_{\langle i,j \rangle} \sum_{\sigma} (\hat{c}_{i,\sigma}^{\dagger} \hat{c}_{j,\sigma} + \hat{c}_{j,\sigma}^{\dagger} \hat{c}_{i,\sigma}) \\
\hat{V}_2^{'} &= \frac{J_H}{2} \sum_{\langle i,j \rangle}(\hat{S}_{i}^{+}\hat{S}_{j}^{-}+\hat{S}_{j}^{-}\hat{S}_{i}^{+})\\
\hat{V}_2^{''} &= J_H \sum_{\langle i,j \rangle}\hat{S}_{i}^{z}\hat{S}_{j}^{z}\\[-1ex]
\end{aligned}
\end{equation}

% Up to second order perturbations, the corrections to unperturbed eigenstate and eigenenergy are

% \begin{equation}
% \ket{\psi_{n}} = \sum_{\alpha} c_{\alpha} (\ket{\psi_{n\alpha}^{(0)}} + \lambda \sum_{m \neq n} \frac{V_{m,n\alpha}}{E_n^{(0)}-E_m^{(0)}} \ket{\psi_{m}^{(0)}}) + O(\lambda^2),\\[-2ex]
% \end{equation}
% \begin{equation}
% E_n - E_n^{(0)} = \lambda \sum_{\alpha,\beta} c_{\alpha}^{*} c_{\beta} V_{n\alpha,n\beta}  + \lambda^2 \sum_{m \neq n} c_{\alpha}^{*} c_{\beta} \frac{ V_{n\alpha,m} V_{m,n\beta}}{E_n^{(0)}-E_m^{(0)}} + O(\lambda^3)
% \end{equation}

% where $V_{n\alpha,n\beta} = \bra{\psi_{n\alpha}^{(0)}} \hat{V} \ket{\psi_{n\beta}^{(0)}}$ and $\ket{\psi_n} = \sum_{\alpha} c_{\alpha} \ket{\psi_{n\alpha}^{(0)}}$.\\

The mapping between these hole operators and the operators in the low-energy Hilbert space can be locally established as
$\hat{h}_{i\updownarrow} \leftrightarrow |\Diamond \rangle \langle \mud{\varnothing}{\updownarrow}|$.
Then, based on the first order perturbation, we derive the effective hopping and Heisenberg exchange between nearest neighboring sites, expressed in terms of hole operators:
\begin{equation}
\begin{aligned}
\bra{\Diamond \mud{\varnothing}{S}} \hat{V}_1 \ket{\mud{\varnothing}{S}\Diamond} = -\frac{t}{2} 
& \quad \longrightarrow \quad \frac{t}{2} \sum_{\langle i,j \rangle} \sum_{\sigma} (1-\hat{n}_{i,\bar{\sigma}}^h) \hat{h}_{i,\sigma}^{\dagger} \hat{h}_{j, \sigma} (1-\hat{n}_{j,\bar{\sigma}}^h) + h.c., \\[-2ex]
\begin{rcases}
\bra{\mud{\varnothing}{\bar{S}} \mud{\varnothing}{S}} \hat{V}_2^{'} \ket{\mud{\varnothing}{S} \mud{\varnothing}{\bar{S}}} = \frac{J_H}{2}\\
\bra{\mud{\varnothing}{S} \mud{\varnothing}{\bar{S}} } \hat{V}_2^{''} \ket{\mud{\varnothing}{S} \mud{\varnothing}{\bar{S}}} = -\frac{J_H}{4}\\
\bra{\mud{\varnothing}{S} \mud{\varnothing}{S} } \hat{V}_2^{''} \ket{\mud{\varnothing}{S} \mud{\varnothing}{S}} = \frac{J_H}{4}
\end{rcases}
& \quad \longrightarrow \quad  J_H\sum_{\langle i,j \rangle} \bm{\hat{S}}_i \cdot \bm{\hat{S}}_j,\\[-1ex]
\end{aligned}
\end{equation}
where $\bm{\hat{S}}_i \cdot \bm{\hat{S}}_j = \frac{1}{2}(\hat{h}_{i,\uparrow}^{\dagger} \hat{h}_{i,\downarrow} \hat{h}_{j,\downarrow}^{\dagger} \hat{h}_{j,\uparrow} + \hat{h}_{i,\downarrow}^{\dagger} \hat{h}_{i,\uparrow}\hat{h}_{j,\uparrow}^{\dagger} \hat{h}_{j,\downarrow}) + \sum_{\sigma,\sigma'}\frac{(-1)^{\sigma\neq \sigma'}}{4} \hat{n}_{i,\sigma}^{h} \hat{n}_{j,\sigma'}^{h}$.\\

The low-energy Hamiltonian due to the correction of the second-order perturbation is written as $\frac{V_{0\alpha,m} V_{m,0\beta}}{E_0^{(0)}-E_m^{(0)}} \ket{\psi_{0\beta}^{(0)}}\bra{\psi_{0\alpha}^{(0)}}$, 
Here, $V_{n\alpha,n\beta} = \bra{\psi_{n\alpha}^{(0)}} \hat{V} \ket{\psi_{n\beta}^{(0)}}$, and $E_n^{(0)}$ is the energy level of $\hat{H_0}$ that has $d$-fold degenerate eigenstates $\ket{\psi_{n\alpha}^{(0)}}$ with $\alpha=1,...,d$.
In our model, the effective operators associated with second-order corrections include:
\begin{equation}
\begin{aligned}
-(\frac{2 t^2}{3 J_K} +\frac{3J_H^2}{32J_K})\ket{\Diamond\Diamond}\bra{\Diamond\Diamond}
\quad \longrightarrow \quad
& -(\frac{2 t^2}{3 J_K} +\frac{3J_H^2}{32J_K}) \sum_{\langle i,j \rangle} (1-\hat{n}_i^h)(1-\hat{n}_j^h), \\
-(\frac{3t^2}{4J_K}+\frac{3J_H^2}{16J_K})\ket{\mud{\varnothing}{S}\Diamond}\bra{\mud{\varnothing}{S}\Diamond} 
\quad \longrightarrow \quad
& -(\frac{3t^2}{4J_K}+\frac{3J_H^2}{16J_K}) \sum_{\langle i,j \rangle} \big(\hat{n}_i^h(1-\hat{n}_j^h) +(1-\hat{n}_i^h)\hat{n}_j^h \big), \\
\frac{3 t J_H}{4 J_K} \ket{\Diamond\mud{\varnothing}{S}}\bra{\mud{\varnothing}{S}\Diamond}
\quad \longrightarrow \quad
& -\frac{3 t J_H}{4 J_K} \sum_{\langle i,j \rangle} \sum_{\sigma} ((1-\hat{n}_{i,\bar{\sigma}}^h)\hat{h}_{i,\sigma}^{\dagger}\hat{h}_{j,\sigma}(1-\hat{n}_{j,\bar{\sigma}}^h) + h.c.),\\
+ \frac{t^2}{6 J_K} \ket{\Diamond\Diamond\mud{\varnothing}{S}}\bra{\mud{\varnothing}{S}\Diamond\Diamond}
\quad \longrightarrow \quad
& + \frac{t^2}{6 J_K}\sum_{\langle i,j,k \rangle} \sum_{\sigma} (\hat{h}_{i,\sigma}^{\dagger}(1-\hat{n}_j^h)\hat{h}_{k,\sigma} +h.c.).
\end{aligned}
\end{equation}

In addition to the items directly associated with an operator mentioned above, there are also second-order corrections that can be attributed to a set of operators:
\begin{equation}
\begin{aligned}
-\frac{t^2}{2J_K} \sum_{S}(\ket{\Diamond \mud{\varnothing}{S} \mud{\varnothing}{\bar{S}}} -\frac{1}{2}\ket{\Diamond \mud{\varnothing}{\bar{S}} \mud{\varnothing}{S}})\bra{\mud{\varnothing}{\bar{S}}\mud{\varnothing}{S}\Diamond}
\quad \longrightarrow \quad
& \frac{t^2}{2J_K} \hat{\xi}_{j,k}^{\dagger} \hat{\xi}_{i,j} + \frac{t^2}{4J_K} \sum_{\sigma} \hat{h}_{k,\sigma}^{\dagger} \hat{n}_{j,\bar{\sigma}}^h  \hat{h}_{i,\sigma} \\
-\frac{t^2}{2J_K} \sum_{S}(\ket{\mud{\varnothing}{\bar{S}} \mud{\varnothing}{S} \Diamond} -\frac{1}{2}\ket{\mud{\varnothing}{S} \mud{\varnothing}{\bar{S}} \Diamond})\bra{\Diamond\mud{\varnothing}{S}\mud{\varnothing}{\bar{S}}}
\quad \longrightarrow \quad
& \frac{t^2}{2J_K} \hat{\xi}_{i,j}^{\dagger} \hat{\xi}_{j,k} + \frac{t^2}{4J_K} \sum_{\sigma} \hat{h}_{i,\sigma}^{\dagger} \hat{n}_{j,\bar{\sigma}}^h  \hat{h}_{k,\sigma}\\[-3ex]
\end{aligned}
\end{equation}

\begin{equation}
\begin{aligned}
\frac{t J_H}{4J_K} \sum_{S} (\ket{\mud{\varnothing}{S} \Diamond \mud{\varnothing}{\bar{S}}} -\frac{1}{2} \ket{\mud{\varnothing}{\bar{S}} \Diamond \mud{\varnothing}{S}})\bra{\mud{\varnothing}{\bar{S}}\mud{\varnothing}{S}\Diamond}
\quad \longrightarrow \quad
& \frac{t J_H}{2J_K} \hat{\xi}_{i,k}^{\dagger} \hat{\xi}_{i,j} - \frac{t J_H}{8 J_K} \sum_{\sigma} \hat{h}_{k,\sigma}^{\dagger} \hat{n}_{i,\bar{\sigma}}^h  \hat{h}_{j,\sigma}\\
\frac{t J_H}{4J_K} \sum_{S} (\ket{\mud{\varnothing}{\bar{S}} \Diamond \mud{\varnothing}{S}} -\frac{1}{2} \ket{\mud{\varnothing}{S} \Diamond \mud{\varnothing}{\bar{S}}})\bra{\Diamond\mud{\varnothing}{S}\mud{\varnothing}{\bar{S}}}
\quad \longrightarrow \quad
& \frac{t J_H}{2J_K} \hat{\xi}_{i,k}^{\dagger} \hat{\xi}_{j,k} - \frac{t J_H}{8 J_K} \sum_{\sigma} \hat{h}_{i,\sigma}^{\dagger} \hat{n}_{k,\bar{\sigma}}^h  \hat{h}_{j,\sigma}\\[-3ex]
\end{aligned}
\end{equation}

\begin{equation}
\begin{aligned}
\frac{t J_H}{8J_K} \sum_{S} \ket{\mud{\varnothing}{S}\Diamond\mud{\varnothing}{S}}\bra{\mud{\varnothing}{S}\mud{\varnothing}{S}\Diamond}
\quad \longrightarrow \quad
& - \frac{t J_H}{8J_K} \sum_{\sigma} \hat{h}_{k,\sigma}^{\dagger} \hat{n}_{i,\sigma}^h  \hat{h}_{j,\sigma} \\
\frac{t J_H}{8J_K} \sum_{S} \ket{\mud{\varnothing}{S}\Diamond\mud{\varnothing}{S}}\bra{\Diamond\mud{\varnothing}{S}\mud{\varnothing}{S}}
\quad \longrightarrow \quad
&- \frac{t J_H}{8J_K} \sum_{\sigma} \hat{h}_{i,\sigma}^{\dagger} \hat{n}_{k,\sigma}^h  \hat{h}_{j,\sigma} \\[-3ex]
\end{aligned}
\end{equation}

\begin{equation}
\begin{aligned}
-\frac{t^2}{4J_K} \sum_{S} \ket{\Diamond\mud{\varnothing}{S}\mud{\varnothing}{S}} \bra{\mud{\varnothing}{S}\mud{\varnothing}{S}\Diamond}
\quad \longrightarrow \quad
& \frac{t^2}{4J_K} \sum_{\sigma} \hat{h}_{k,\sigma}^{\dagger} \hat{n}_{j,\sigma}^h  \hat{h}_{i,\sigma}\\
-\frac{t^2}{4J_K} \sum_{S} \ket{\Diamond\mud{\varnothing}{S}\mud{\varnothing}{S}} \bra{\Diamond\mud{\varnothing}{S}\mud{\varnothing}{S}}
\quad \longrightarrow \quad
& \frac{t^2}{4J_K} \sum_{\sigma} \hat{h}_{i,\sigma}^{\dagger} \hat{n}_{j,\sigma}^h  \hat{h}_{k,\sigma}\\[-2ex]
\end{aligned}
\end{equation}

\begin{equation}
\begin{aligned}
\frac{t J_H}{8 J_K} \sum_{S} (\ket{\mud{\varnothing}{S} \mud{\varnothing}{S} \Diamond}+ \ket{\Diamond \mud{\varnothing}{S} \mud{\varnothing}{S}}) \bra{\mud{\varnothing}{S} \Diamond \mud{\varnothing}{S}}
\quad \longrightarrow \quad
& - \frac{t J_H}{8 J_K} \sum_{\sigma} \hat{h}_{j,\sigma}^{\dagger} (\hat{n}_{i,\sigma}^h  \hat{h}_{k,\sigma} + \hat{n}_{k,\sigma}^h  \hat{h}_{i,\sigma}) \\
-\frac{J_H^2}{8 J_K} \sum_{S} \ket{\mud{\varnothing}{S} \Diamond \mud{\varnothing}{S}}\bra{\mud{\varnothing}{S} \Diamond \mud{\varnothing}{S}}
\quad \longrightarrow \quad
& -\frac{J_H^2}{8 J_K} \sum_{\sigma} \hat{n}_{i,\sigma}^h \hat{n}_{k,\sigma}^h (1-\hat{n}_j^h)\\[-3ex]
\end{aligned}
\end{equation}

\begin{equation}
\begin{aligned}
\shortstack{$
\begin{aligned}
\sum_{S} \Big(\frac{t J_H}{4 J_K} & (\ket{\mud{\varnothing}{\bar{S}} \mud{\varnothing}{S} \Diamond} + \ket{\Diamond \mud{\varnothing}{\bar{S}} \mud{\varnothing}{S}} \\[-2ex]
& -\frac{1}{2}\ket{\mud{\varnothing}{S} \mud{\varnothing}{\bar{S}} \Diamond} -\frac{1}{2} \ket{\Diamond \mud{\varnothing}{S} \mud{\varnothing}{\bar{S}}})\Big)\bra{\mud{\varnothing}{S}\Diamond\mud{\varnothing}{\bar{S}}}
\end{aligned}
$}
\quad \longrightarrow \quad
& \frac{t J_H}{2 J_K} (\hat{\xi}_{j,k}^{\dagger} \hat{\xi}_{i,k} + \hat{\xi}_{i,j}^{\dagger} \hat{\xi}_{i,k}) - \frac{t J_H}{8 J_K} ( \sum_{\sigma} \hat{h}_{j,\sigma}^{\dagger} \hat{n}_{k,\bar{\sigma}}^h  \hat{h}_{i,\sigma} + \sum_{\sigma} \hat{h}_{j,\sigma}^{\dagger} \hat{n}_{i,\bar{\sigma}}^h  \hat{h}_{k,\sigma}) \\
-\frac{J_H^2}{4 J_K} \sum_{S} (\ket{\mud{\varnothing}{\bar{S}} \Diamond \mud{\varnothing}{S}}-\frac{1}{2}\ket{\mud{\varnothing}{S} \Diamond \mud{\varnothing}{\bar{S}}})\bra{\mud{\varnothing}{S}\Diamond\mud{\varnothing}{\bar{S}}}
\quad \longrightarrow \quad
& -\frac{J_H^2}{4 J_K} \sum_{\sigma} \hat{S}_i^{\bar{\sigma}} \hat{S}_k^\sigma (1-\hat{n}_j^h) + \frac{J_H^2}{8 J_K} \sum_{\sigma} \hat{n}_{i,\sigma}^h \hat{n}_{k,\bar{\sigma}}^h \\[-2ex]
\end{aligned}
\end{equation}
Note that  $\hat{S}_i^{\sigma} \hat{S}_j^{\bar{\sigma}} = \hat{h}_{i,\sigma}^{\dagger}\hat{h}_{i,\bar{\sigma}}\hat{h}_{j,\bar{\sigma}}^{\dagger}\hat{h}_{j,\sigma}$.\\

By aggregating all the first- and second-order corrections, we derive the effective Hamiltonian in real space:
\begin{equation}
\begin{aligned}
\hat{H} = &+J_H\sum_{\langle i,j \rangle} \bm{\hat{S}}_i \cdot \bm{\hat{S}}_j -\frac{J_H^2}{2 J_K} \sum_{\langle i,j,k\rangle} \bm{\hat{S}_i}\cdot\bm{\hat{S}_k}(1-\hat{n}_j^h)\\
& + (\frac{t}{2} - \frac{3 t J_H}{4 J_K}) \sum_{\langle i,j \rangle} \sum_{\sigma} (1-\hat{n}_{i,\bar{\sigma}}^h) \hat{h}_{i,\sigma}^{\dagger} \hat{h}_{j, \sigma} (1-\hat{n}_{j,\bar{\sigma}}^h) + h.c.\\
& -(\frac{2 t^2}{3 J_K} +\frac{3J_H^2}{32J_K}) \sum_{\langle i,j \rangle} (1-\hat{n}_i^h)(1-\hat{n}_j^h)
-(\frac{3t^2}{4J_K}+\frac{3J_H^2}{16J_K}) \sum_{\langle i,j \rangle} \big(\hat{n}_i^h(1-\hat{n}_j^h) +(1-\hat{n}_i^h)\hat{n}_j^h \big) \\ 
& +\frac{t^2}{6 J_K}\sum_{\langle i,j,k \rangle} \sum_{\sigma} ((1-\hat{n}_{i,\bar{\sigma}}^h) \hat{h}_{i,\sigma}^{\dagger}(1-\hat{n}_j^h)\hat{h}_{k,\sigma}  (1-\hat{n}_{k,\bar{\sigma}}^h) +h.c.) \\
&+ \frac{t^2}{4J_K} \sum_{\langle i,j,k\rangle} \sum_{\sigma} (\hat{h}_{i,\sigma}^{\dagger} \hat{n}_{j}^h  \hat{h}_{k,\sigma} + h.c.) 
- \frac{t J_H}{8 J_K} \sum_{\langle i,j,k\rangle} \sum_{\sigma} \big((\hat{h}_{i,\sigma}^{\dagger} \hat{n}_{k}^h + \hat{h}_{k,\sigma}^{\dagger} \hat{n}_{i}^h)  \hat{h}_{j,\sigma} +h.c.\big) \\
& + \frac{t^2}{2J_K} \sum_{\langle i,j,k\rangle} (\hat{\xi}_{i,j}^{\dagger} \hat{\xi}_{j,k} +h.c.)  + \frac{t J_H}{2J_K} \sum_{\langle i,j,k\rangle} \big((\hat{\xi}_{i,j}^{\dagger} +\hat{\xi}_{j,k}^{\dagger})\hat{\xi}_{i,k} + h.c.\big) \\
\end{aligned}
\end{equation}

\section{Effective Hamiltonian}

Transforming the Hamiltonian on the square lattice into momentum space, we recast it into the following form:
\begin{equation}\label{eq:effectiveH}
\begin{aligned}
\hat{H} \; &= \hat{H}_1 + \hat{H}_2 + \hat{H}_3 \\
\hat{H}_1 &= \sum_{\bm{k}}\ \varepsilon(\bm{k}) \hat{n}_{\bm{k}}^{h}\\
\hat{H}_2 &= \frac{1}{N}\sum_{\bm{q},\bm{k},\bm{k}'} \Big(\Gamma_2^{s}(\bm{q};\bm{k},\bm{k}') + \Gamma_2^{t}(\bm{q};\bm{k},\bm{k}')\Big) h^{\dagger}_{\frac{\bm{q}}{2}+\bm{k},\uparrow} h^{\dagger}_{\frac{\bm{q}}{2}-\bm{k},\downarrow} \hat{h}_{\frac{\bm{q}}{2}-\bm{k}',\downarrow} \hat{h}_{\frac{\bm{q}}{2}+\bm{k}',\uparrow}\\
\hat{H}_3 &= \frac{1}{N}\sum_{\bm{q},\bm{k},\bm{k}'}\sum_{\sigma} \Big(\Gamma_3^{s}(\bm{q};\bm{k},\bm{k}') + \Gamma_3^{t}(\bm{q};\bm{k},\bm{k}')\Big) h^{\dagger}_{\frac{\bm{q}}{2}+\bm{k},\sigma} h^{\dagger}_{\frac{\bm{q}}{2}-\bm{k},\sigma} \hat{h}_{\frac{\bm{q}}{2}-\bm{k}',\sigma} \hat{h}_{\frac{\bm{q}}{2}+\bm{k}',\sigma}\\
\end{aligned}
\end{equation}

To investigate different channels, we separate the $S^z=0$ and $S^z=\pm 1$ sectors of the Hamiltonian into symmetric and anti-symmetric parts, where symmetric part represents the singlet channels and the anti-symmetric parts represents the triplet channels. Due to the spin rotation symmetry, we can only focus on $\hat{H}_1$ and $\hat{H}_2$, since the other two triplet states are degenerate with the one with $S^z =0$.

The band dispersion and interaction vertices in Eqn.~\ref{eq:effectiveH} calculated through strong-coupling expansion are 
\begin{equation}
\begin{aligned}
\varepsilon(\bm{k}) &= 
-(\frac{t^2}{6 J_K}+\frac{3 J_H^2}{16 J_K}) 
+ (t - \frac{3 t J_H}{2 J_K})\sum_{\alpha} \cos{k_{\alpha}} 
+ \frac{2 t^2}{3 J_K} \big( (\sum_\alpha \cos k_\alpha )^2 -1 \big)\\
%% \Gamma_2^{s}
\Gamma_2^{s}(\bm{q};\bm{k},\bm{k}') &= 
(-\frac{3 J_H}{2} + \frac{9 J_H^2}{16 J_K} - \frac{2 t^2}{3 J_K}) \sum_{\alpha} \cos{k_\alpha}\cos{k'_\alpha} 
+ \frac{3 J_H^2}{2 J_K} \big( (\sum_{\alpha} \cos{k_{\alpha}}\cos{k'_{\alpha}})^2+ (\sum_{\alpha} \sin{k_{\alpha}}\sin{k'_{\alpha}})^2 - 1 \big)\\
&\quad 
- (2t - \frac{3 t J_H}{2 J_K}) \sum_{\alpha} \cos{(\frac{q_\alpha}{2})} (\cos{k_\alpha}+\cos{k'_\alpha})
+ \frac{3 t J_H}{2 J_K} \sum_{\alpha,\beta} \cos{(\frac{q_\alpha}{2})} (\cos{k_\alpha}+\cos{k'_\alpha}) \cos{k_\beta}\cos{k'_\beta} \\
&\quad 
-\frac{4 t^2}{3 J_K} \big( (\sum_{\alpha} \cos{\frac{q_{\alpha}}{2}} \cos{k_{\alpha}})^2 + (\sum_{\alpha} \cos{\frac{q_{\alpha}}{2}} \cos{k'_{\alpha}})^2 + (\sum_{\alpha} \sin{\frac{q_{\alpha}}{2}} \sin{k_{\alpha}})^2 + (\sum_{\alpha} \sin{\frac{q_{\alpha}}{2}} \sin{k'_{\alpha}})^2 -2 \big)\\
&\quad 
+ \frac{14 t^2}{3 J_K} \sum_{\alpha,\beta} \cos{(\frac{q_\alpha}{2})}\cos{k_\alpha}\cos{(\frac{q_\beta}{2})}\cos{k_\beta'}\\
%% \Gamma_2^{t}
\Gamma_2^{t}(\bm{q};\bm{k},\bm{k}') &= 
(\frac{J_H}{2} + \frac{9 J_H^2}{16 J_K} + \frac{4 t^2}{3 J_K}) \sum_{\alpha} \sin{k_\alpha}\sin{k'_\alpha} 
- \frac{J_H^2}{J_K} \sum_{\alpha,\beta} \cos{k_{\alpha}} \cos{k'_{\alpha}} \sin{k_{\beta}} \sin{k'_{\beta}} \\
&\quad 
- \frac{t J_H}{J_K} \sum_{\alpha,\beta} \cos{(\frac{q_\alpha}{2})} (\cos{k_\alpha}+\cos{k'_\alpha}) \sin{k_\beta}\sin{k'_\beta}  +  \frac{2 t^2}{3 J_K} \sum_{\alpha,\beta} \sin{(\frac{q_\alpha}{2})} \sin{k_{\alpha}} \sin{(\frac{q_\beta}{2})} \sin{k'_{\beta}} \\
% %% \Gamma_3^{s}
% \Gamma_3^{s}(\bm{q};\bm{k},\bm{k}') &= 
% (\frac{J_H}{4} + \frac{9 J_H^2}{32 J_K} + \frac{2 t^2}{3 J_K}) \sum_{\alpha} \cos{k_\alpha}\cos{k'_\alpha}
%  + \frac{t J_H}{4 J_K} \sum_{\alpha,\beta} \cos{(\frac{q_\alpha}{2})} (\cos{k_\alpha}+\cos{k'_\alpha}) \\
% &\quad 
% - \frac{J_H^2}{8 J_K} \sum_{\alpha} \cos{2 k_\alpha}\cos{2 k'_\alpha} 
% - \frac{J_H^2}{4 J_K} (\cos{k_x}\cos{k_x'}\cos{k_y}\cos{k_y'} + \sin{k_x}\sin{k_x'}\sin{k_y}\sin{k_y'})\\
% &\quad 
% - \frac{t J_H}{2 J_K} \sum_{\alpha,\beta} \cos{(\frac{q_\alpha}{2})} (\cos{k_\alpha}+\cos{k'_\alpha}) \cos{k_\beta}\cos{k'_\beta} 
% + \frac{t^2}{3 J_K} \sum_{\alpha,\beta} \cos{(\frac{q_\alpha}{2})}\cos{k_{\alpha}} \cos{(\frac{q_\beta}{2})} \cos{k'_{\beta}} \\
% %% \Gamma_3^{t} %% f3 antisymmetric
% \Gamma_3^{t}(\bm{q};\bm{k},\bm{k}') &= \frac{1}{2} \Gamma_2^{t}(\bm{q};\bm{k},\bm{k}')\\
\end{aligned}
\end{equation}

\section{Mean Field calculation}

Define the superconducting order parameter as
\begin{equation}
\phi_{\bm{k}}^{(\bm{q})} = \langle \hat{h}_{\frac{\bm{q}}{2}-\bm{k},\downarrow} \hat{h}_{\frac{\bm{q}}{2}+\bm{k},\uparrow} \rangle_{\text{MF}}, 
\end{equation}
then the singlet and triplet superconducting gap function can be expressed as
\begin{equation}
\begin{aligned}
\Delta_{s}(\bm{q};\bm{k}) &= \frac{1}{N}\sum_{\bm{k}'} \Gamma_2^{s}(\bm{q};\bm{k},\bm{k}') \phi_{\bm{k'}}^{(\bm{q})}; \; \; \; \Delta_{s}(\bm{q}) = \sqrt{\frac{1}{N}\sum_{\bm{k}} \Delta_{s}^{*}(\bm{q};\bm{k}) \Delta_{s}(\bm{q};\bm{k})}; \\
\Delta_{t}(\bm{q};\bm{k}) &= \frac{1}{N}\sum_{\bm{k}'} \Gamma_2^{t}(\bm{q};\bm{k},\bm{k}') \phi_{\bm{k'}}^{(\bm{q})}; \; \; \; \Delta_{t}(\bm{q}) = \sqrt{\frac{1}{N}\sum_{\bm{k}} \Delta_{t}^{*}(\bm{q};\bm{k}) \Delta_{t}(\bm{q};\bm{k})}; \\
\end{aligned}
\end{equation}

% Note that $\Delta_{t}(\bm{q};\bm{k})$ degenerates with $\sum_{\bm{k}'} \Gamma_3^{t}(\bm{q};\bm{k},\bm{k}') \phi_{\sigma\sigma}(\bm{q};\bm{k}')$ due to SU(2) symmetry, and $\sum_{\bm{k}'} \Gamma_3^{s}(\bm{q};\bm{k},\bm{k}') \phi_{\sigma\sigma}(\bm{q};\bm{k}')=0$ due to its anti-symmetry. 
We perform mean-field approach exclusively to $\hat{H}_1$ and $\hat{H}_2$, yielding the mean-field Hamiltonian as
\begin{equation}
\begin{aligned}
\hat{H}_{\text{MF}} = \sum_{\bm{k}}\ \varepsilon(\bm{k}) \hat{n}_{\bm{k}}^{h} \quad 
&
+ \quad \sum_{\bm{q},\bm{k}} \Big((\Delta_s(\bm{q};\bm{k}) + \Delta_t(\bm{q};\bm{k})) h^{\dagger}_{\frac{\bm{q}}{2}+\bm{k},\uparrow} h^{\dagger}_{\frac{\bm{q}}{2}-\bm{k},\downarrow} \\
& \quad \quad \quad \quad 
+ (\Delta^*_s(\bm{q};\bm{k}) + \Delta^*_t(\bm{q};\bm{k})) \hat{h}_{\frac{\bm{q}}{2}-\bm{k},\downarrow} \hat{h}_{\frac{\bm{q}}{2}+\bm{k},\uparrow} \\
& \quad \quad \quad \quad 
- (\Delta^*_s(\bm{q};\bm{k}) + \Delta^*_t(\bm{q};\bm{k})) \phi_{\bm{k}}^{(\bm{q})}
\Big).
\end{aligned}
\end{equation}

Here, we take the mean-field ansatz where the Cooper pair can take only a fixed value of total momentum. It leads to a Schrödinger-like equation for each total momentum $\bm{q}$ and relative momentum $\bm{k}$:
\begin{equation}\label{eq:H_matrix}
\begin{pmatrix}
\varepsilon(\frac{\bm{q}}{2}+\bm{k}) -\mu & \Delta_s(\bm{q};\bm{k}) + \Delta_t(\bm{q};\bm{k})\\
\Delta^*_s(\bm{q};\bm{k}) + \Delta^*_t(\bm{q};\bm{k}) & -\varepsilon(\frac{\bm{q}}{2}-\bm{k}) + \mu
\end{pmatrix}
\begin{pmatrix}
\hat{h}_{\frac{\bm{q}}{2}+\bm{k},\uparrow}\\
h^{\dagger}_{\frac{\bm{q}}{2}-\bm{k},\downarrow}
\end{pmatrix}
= E(\bm{q};\bm{k})
\begin{pmatrix}
\hat{h}_{\frac{\bm{q}}{2}+\bm{k},\uparrow}\\
h^{\dagger}_{\frac{\bm{q}}{2}-\bm{k},\downarrow}.
\end{pmatrix}
\end{equation}

To further simplify the solution, we introduce variables
\begin{equation}
\begin{aligned}
& \bar{\varepsilon} = \frac{1}{2} \big( (\varepsilon(\frac{\bm{q}}{2}+\bm{k})-\mu) +(\varepsilon(\frac{\bm{q}}{2}-\bm{k}) -\mu)\big), \\
& \delta=\frac{1}{2} \big((\varepsilon(\frac{\bm{q}}{2}+\bm{k}) -\mu) - (\varepsilon(\frac{\bm{q}}{2}-\bm{k})-\mu) \big), \\
& \Delta =\Delta_s(\bm{q};\bm{k}) + \Delta_t(\bm{q};\bm{k}), \\
\end{aligned}
\end{equation}
then Eqn.~\ref{eq:H_matrix} is simplified to be
\begin{equation}
\begin{pmatrix}
\bar{\varepsilon} + \delta  & \Delta\\
\Delta^* & -\bar{\varepsilon} + \delta
\end{pmatrix}
\begin{pmatrix}
\hat{h}_{\frac{\bm{q}}{2}+\bm{k},\uparrow}\\
h^{\dagger}_{\frac{\bm{q}}{2}-\bm{k},\downarrow}
\end{pmatrix}
= E(\bm{q};\bm{k})
\begin{pmatrix}
\hat{h}_{\frac{\bm{q}}{2}+\bm{k},\uparrow}\\
h^{\dagger}_{\frac{\bm{q}}{2}-\bm{k},\downarrow}
\end{pmatrix}.
\end{equation}

The eigenvalues and the corresponding eigenvectors of the above matrix can be expressed in terms of these parameters:
\begin{equation}
\begin{aligned}
\epsilon_\pm (\bm{q};\bm{k}) &= \delta \pm \sqrt{\Delta^2+\bar{\varepsilon}^2} \\
\bm{v}_{\pm} (\bm{q};\bm{k}) &= \frac{\bar{\varepsilon} \pm \sqrt{\Delta^2+\bar{\varepsilon}^2}}{\sqrt{\Delta^2 + (\bar{\varepsilon} \pm \sqrt{\Delta^2+\bar{\varepsilon}^2})^2}} \hat{h}_{\frac{\bm{q}}{2}+\bm{k},\uparrow} 
+\frac{\Delta}{\sqrt{\Delta^2 + (\bar{\varepsilon} \pm \sqrt{\Delta^2+\bar{\varepsilon}^2})^2}} h^{\dagger}_{\frac{\bm{q}}{2}-\bm{k},\downarrow} \\
\end{aligned}
\end{equation}

Then, we can obtain the superconducting order parameter and the hole density:
\begin{equation}
\phi_{\bm{k}}^{(\bm{q})} = \langle \hat{h}_{\frac{\bm{q}}{2}-\bm{k},\downarrow} \hat{h}_{\frac{\bm{q}}{2}+\bm{k},\uparrow} \rangle _{\text{MF}} 
= \frac{\Delta}{2\sqrt{\Delta^2+\bar{\varepsilon}^2}} (\frac{1}{e^{\beta \epsilon_{+}}+1}-\frac{1}{e^{\beta \epsilon_{-}}+1})\\[-2ex]
\end{equation}
\begin{equation}
n^{(\bm{q})} = \frac{1}{N}\sum_{\bm{k}} \langle \hat{h}_{\frac{\bm{q}}{2}+\bm{k},\uparrow}^{\dagger} \hat{h}_{\frac{\bm{q}}{2}+\bm{k},\uparrow} \rangle + \langle \hat{h}_{\frac{\bm{q}}{2}-\bm{k},\downarrow}^{\dagger} \hat{h}_{\frac{\bm{q}}{2}-\bm{k},\downarrow} \rangle 
= \frac{1}{N} \sum_{\bm{k}}\big( 1 + \frac{\bar{\varepsilon}}{\sqrt{\Delta^2+\bar{\varepsilon}^2}} (\frac{1}{e^{\beta \epsilon_{+}}+1} - \frac{1}{e^{\beta \epsilon_{-}}+1})\big)
\end{equation}

The resulting mean-field grand potential is
\begin{equation}
\begin{aligned}
G &= -\frac{1}{\beta} \ln{Z} = -\frac{1}{\beta} \ln{\Tr(e^{-\beta \hat{H}})}
= -\frac{1}{\beta} \ln{\Tr(e^{-\beta (\hat{H}-\hat{H}_{tr})}e^{-\beta \hat{H}_{tr}})} \\
& = -\frac{1}{\beta} \big(\ln{\frac{\Tr(e^{-\beta (\hat{H}-\hat{H}_{tr})}e^{-\beta \hat{H}_{tr}})}{\Tr(e^{-\beta \hat{H}_{tr}})}} + \ln{\Tr(e^{-\beta \hat{H}_{tr}})} \big) =\ln{ \langle e^{-\beta (\hat{H}-\hat{H}_{tr})} \rangle_{tr}} + G_{tr}\\
& \leq -\frac{1}{\beta} \ln{e^{-\beta \langle \hat{H}-\hat{H}_{tr} \rangle_{tr}}} + G_{tr} = \langle \hat{H}-\hat{H}_{tr} \rangle_{tr} + G_{tr}.\\
\end{aligned}
\end{equation}

The grand potential of trial Hamiltonian is $G_{tr}^{(\bm{q})} = -\frac{1}{\beta} \ln{\prod_{\bm{k},i}(1+e^{-\beta \epsilon_i(\bm{q};\bm{k})})}$, and $\langle \hat{H}-\hat{H}_{tr} \rangle_{tr} \approx - \sum_{\bm{k}}(\Delta^*_s(\bm{q};\bm{k}) + \Delta^*_t(\bm{q};\bm{k})) \phi_{\bm{k}}^{(\bm{q})}$. As the grand potential is related to Helmholtz free energy by $F_H = G + \mu N$,  the mean-field Helmholtz free energy at each $\bm{q}$ is
\begin{equation}
F_H^{(\bm{q})} = -\frac{1}{\beta} \ln{\prod_{\bm{k},i}(1+e^{-\beta \epsilon_i(\bm{q};\bm{k})})}  - \sum_{\bm{k}}(\Delta^*_s(\bm{q};\bm{k}) + \Delta^*_t(\bm{q};\bm{k})) \phi_{\bm{k}}^{(\bm{q})} + \mu N.
\end{equation}

In our mean-field calculation, we vary the chemical potential $\mu$ to keep the hole density $n^{(\bm{q})}$ a fixed value. While $n^{(\bm{q})}$ is used for self-consistency, we update $\phi_{\bm{k}}^{(\bm{q})}$ by minimizing the Helmholtz free energy $F_H^{(\bm{q})}$.\\

To calculate the condensation energy gain, we calculate the hole density and the Helmholtz free energy of the metallic state, where the interacting terms are set to be zero:
\begin{equation}
\begin{aligned}
&n^0 = \frac{1}{N}\sum_{\bm{k}} \frac{2}{e^{\beta (\varepsilon(\bm{k})-\mu)}+1}\\
&F_{H}^{0} = -\frac{1}{\beta} \ln{\prod_{\bm{k}}(1+e^{-\beta (\varepsilon(\bm{k})-\mu)})^2}.
\end{aligned}
\end{equation}

\section{Symmetries and Order parameters}

Nematic order parameter at $\bm{q}=(0,0)$:
\begin{equation}
\begin{aligned}
O_{\mathcal{N}} &= 1- \frac{| \sum_{\bm{k}} \phi_{R_{\frac{\pi}{2}}\bm{k}}^{(0,0) *} \phi_{\bm{k}}^{(0,0)}|}{| \sum_{\bm{k}} \phi_{\bm{k}}^{(0,0) *} \phi_{\bm{k}}^{(0,0)} |}
\end{aligned}
\end{equation}

Order parameter that detects time-reversal at $\bm{q}=(0,0)$
\begin{equation}
\begin{aligned}
O_{\mathcal{T}} &= 1- \frac{| \sum_{\bm{k}} \phi_{\bm{k}}^{(0,0)} \phi_{\bm{k}}^{(0,0)}|}{| \sum_{\bm{k}} \phi_{\bm{k}}^{(0,0) *} \phi_{\bm{k}}^{(0,0)} |}
\end{aligned}
\end{equation}

The system we are studying possesses the D4 point group symmetry, and the associated order parameters can be expressed as
\begin{equation}
\begin{aligned}
O^{\ell}(\bm{q}) &= \frac{1}{N} \sum_{\bm{k}} f^{\ell}(\bm{q},\bm{k}) \phi_{\bm{k}}^{\bm{(q)}}
\end{aligned}
\end{equation}

The form factors correspond to various symmetry operations are given in the following table:
\begin{equation}
\begin{aligned}
  \begin{array}{c|c|c|c}
    \hline
    \ell & f^{\ell}(\bm{q},\bm{k}) & f^{\ell}((0,0),\bm{k}) & f^{\ell}((\pi,\pi),\bm{k}) \\
    \hline
    s & \cos{(\frac{q_x}{2}-k_x)}+\cos{(\frac{q_y}{2}-k_y)} & \cos{(k_x)}+\cos{(k_y)} & \sin{(k_x)}+\sin{(k_y)} \\
    \hline
    d_{x^2-y^2} & \cos{(\frac{q_x}{2}-k_x)}-\cos{(\frac{q_y}{2}-k_y)} & \cos{(k_x)}-\cos{(k_y)} & \sin{(k_x)}-\sin{(k_y)} \\
    \hline
    d_{xy} & \sin{(\frac{q_x}{2}-k_x)}\sin{(\frac{q_y}{2}-k_y)} & \sin{(k_x)}\sin{(k_y)} & \cos{(k_x)}\cos{(k_y)} \\
    \hline
    p_x & \sin{(\frac{q_x}{2}-k_x)} & -\sin{(k_x)} & \cos{(k_x)}\\
    \hline
    p_y & \sin{(\frac{q_y}{2}-k_y)} & -\sin{(k_y)} & \cos{(k_y)}\\
    \hline
  \end{array}
\end{aligned}
\end{equation}

While the systems admit certain point group $G$, the form of the superconducting order parameter, $\phi_{\bm{k}}^{(\bm{q})}$, must transform as an irrep. of the symmetry group, i.e., 
\begin{align}
R_{ab}(g) \phi_{\bm{k}}^{(\bm{q}) b} = \phi_{g\bm{k}}^{(g\bm{q}) a}, \;\;\;  \forall g \in G
\end{align}
where $R(g)$ is the irrep. of the point group, and $a$ and $b$ label the possible degeneracy.\\

In the restricted Hartree-Fock approach we adopted, for each candidate state, $\phi_{\bm{k}}^{(\bm{q})}$ is only valued on a certain $\bm{q} = \bm{q}_0$. In other words,
\begin{align}
    \phi_{\bm{k}}^{(\bm{q})} \sim \delta(\bm{q}-\bm{q}_0)\varphi_{\bm{k}}.
\end{align}
\\
Therefore, unless $g \bm{q}_0 = \bm{q}_0$ for $\forall g \in G$,  we cannot obtain a meaningful point-group classification of the order parameter, since we are not allowing superposition on various symmetry-equivalent $\bm{q}$'s. It means that, in the cases we studied, $\bm{q}$ must be $(0,0)$ or $(\pi, \pi)$ in 2D, and $0$ or $\pi$ in 1D.\\

\section{Illustration of Superconducting Gap Functions}

As the supplements of the plots of condensation energy in the main text, here we plot the singlet superconducting gap functions for our mean-field effective model on square lattice with hole density $n^{(\bm{q})}=0.125$.

%% Plot

\begin{figure*}[htb!]
\includegraphics[width=0.82\linewidth]{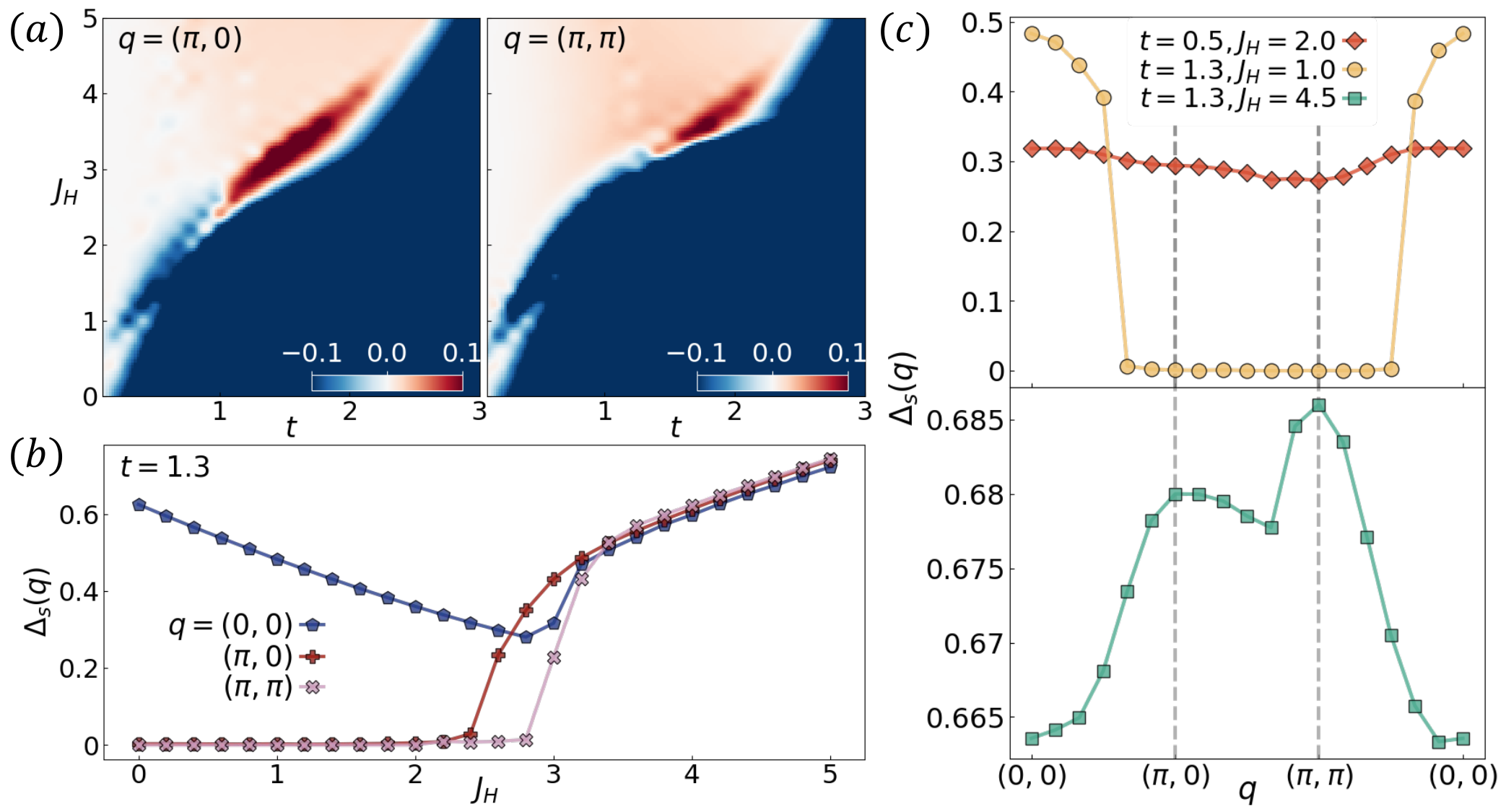}
\caption{\label{2D_Ds} \textbf{Singlet superconducting gap functions in 2D lattice.} (a) $\Delta_s(\bm{q})-\Delta_s(0,0)$ with $\bm{q}=(\pi,0)$ and $(\pi,\pi)$. Example of $\Delta_s(\bm{q})$ are depicted in (b) with $t=1.3$ and in (c) at the scatter points chosen to represent various regions.
Simulations are done with a $12\times 12$ square lattice for (a) and (b), and a $24\times 24$ square lattice for (c).}
\end{figure*}

\begin{figure*}[htb!]
\includegraphics[width=0.82\linewidth]{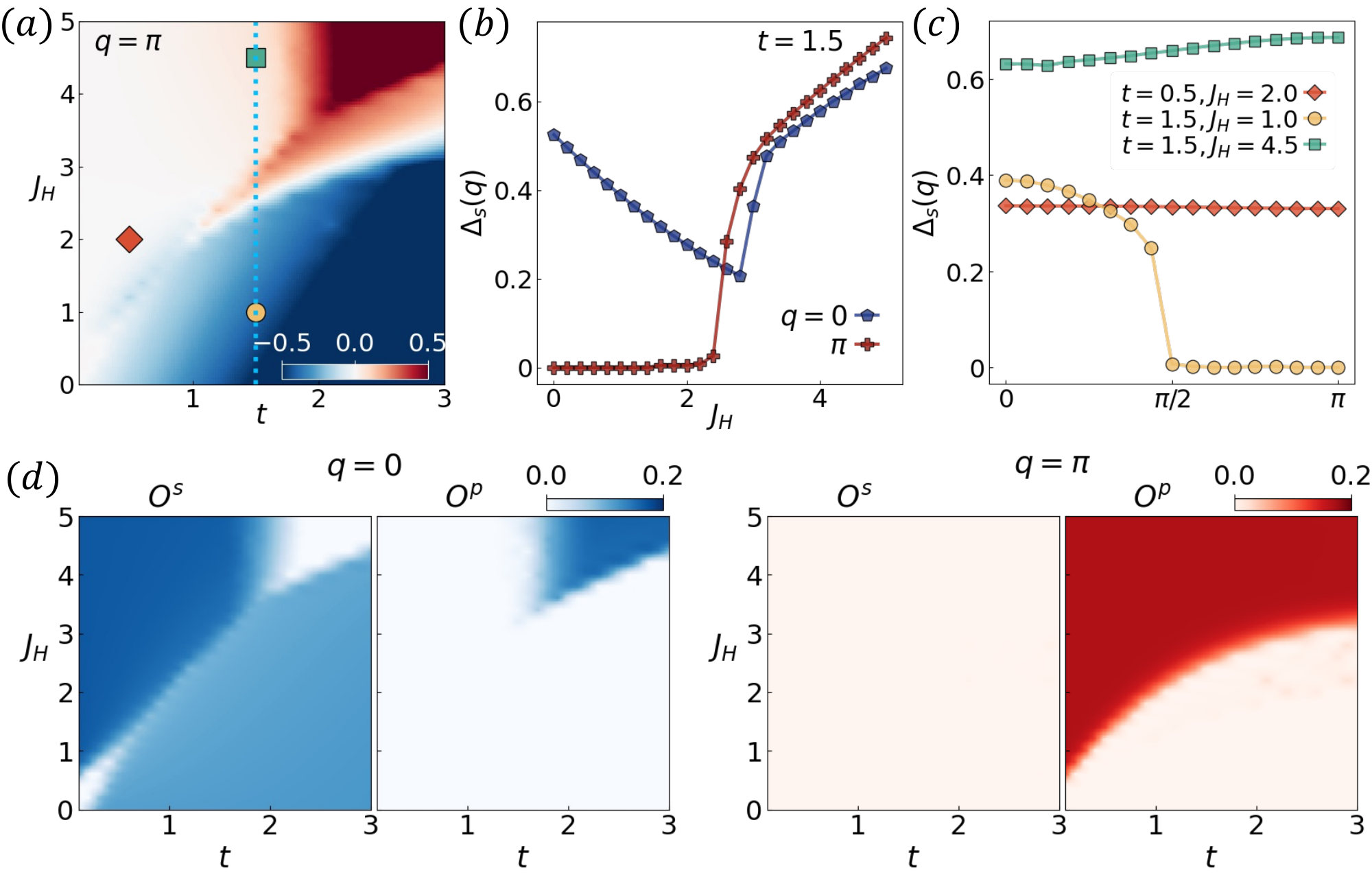}
\caption{\label{1D_Ds} \textbf{Singlet superconducting gap functions in 1D lattice.} (a) $\Delta_s(\pi)-\Delta_s(0)$. Example of $\Delta_s(\bm{q})$ are depicted in (b) along the vertical dotted cut and in (c) at the scatter points chosen to represent various regions. (d) Pairing symmetries of the superconducting order parameters at $q=0$ and $q=\pi$.
Simulations are done with a $N=128$ chain for (a), (b) and (d), and a $N=256$ chain for (c).}
\end{figure*}

% \section{Point group classification of the order parameter}

% In general, the superconducting order parameter(s) in a system (in the $S^z=0$ sector) can be written as
% \begin{align}
%     \Phi^a(\bm{q},\bm{k}) \equiv \langle \hat{c}_{\frac{\bm{q}}{2}+\bm{k},\uparrow} \hat{c}_{\frac{\bm{q}}{2}-\bm{k},\downarrow}\rangle_a
% \end{align}
% where $a$ labels the possible degeneracy.

% Since the systems we are studying admit certain point group $G$, the form of the order parameter, $\Phi(\bm{q},\bm{k})$, must transform as an irrep. of the symmetry group, i.e. $\forall g \in G$
% \begin{align}
% R_{ab}(g) \Phi^b(\bm{q},\bm{k}) = \Phi^a(g\bm{q},g\bm{k})
% \end{align}
% where $R(g)$ is the irrep. of the point group.

% In the restricted Hartree-Fock approach we adopted, for each candidate state, $\Phi(\bm{q},\bm{k})$ is only valued on a certain $\bm{q} = \bm{q}_0$. In other words,
% \begin{align}
%     \Phi(\bm{q},\bm{k}) \sim \delta(\bm{q} - \bm{q_0})\varphi(\bm{k})
% \end{align}
% Therefore, unless $g\bm{q}_0 = \bm{q}_0$ for $\forall g \in G$,  we cannot obtain a meaningful point-group classification of the order parameter, since we are not allowing superposition on various symmetry-equivalent $\bm{q}$'s. In the cases we studied, this means $\bm{q}$ must be $(0,0)$ or $(\pi, \pi)$ in two dimensions, and $0$ or $\pi$ in one dimension.